\documentclass[10pt,letterpaper]{article}
\usepackage[top=0.85in,footskip=0.75in,marginparwidth=2in]{geometry}

\usepackage[utf8]{inputenc}

\usepackage{cite}[sort]

\usepackage{nameref,hyperref}

\usepackage{microtype}
\DisableLigatures[f]{encoding = *, family = * }

\raggedright
\usepackage{changepage}

\usepackage[aboveskip=1pt,labelfont=bf,labelsep=period,singlelinecheck=off]{caption}

\makeatletter
\renewcommand{\@biblabel}[1]{\quad#1.}
\makeatother

\usepackage{lastpage,fancyhdr,graphicx}
\usepackage{epstopdf}
\fancyheadoffset[L]{2.25in}
\fancyfootoffset[L]{2.25in}

\usepackage{color}

\definecolor{Gray}{gray}{.25}

\usepackage{graphicx}

\usepackage{sidecap}

\usepackage{wrapfig}
\usepackage[pscoord]{eso-pic}
\usepackage[fulladjust]{marginnote}
\reversemarginpar

\begin{document}

\begin{flushleft}
{\Large
\textbf\newline{Low Latency Computing for Time Stretch Instruments}
}
\newline
\\
Tingyi Zhou\textsuperscript{1},
Bahram Jalali\textsuperscript{1,*},
\bigskip
\bf{1} Department of Electrical and Computer Engineering, University of California, Los Angeles
\\
\bigskip
*jalali@ucla.edu

\end{flushleft}

\section*{Abstract}
Time stretch instruments have been exceptionally successful in discovering single-shot ultrafast phenomena such as optical rogue waves and have led to record-speed microscopy, spectroscopy, lidar, etc. These instruments encode the ultrafast events into the spectrum of a femtosecond pulse and then dilate the time scale of the data using group velocity dispersion. Generating as much as Tbit per second of data, they are ideal partners for deep learning networks which by their inherent complexity, require large datasets for training. However, the inference time scale of neural networks in the millisecond regime is orders of magnitude longer than the data acquisition rate of time stretch instruments. This underscores the need to explore means where some of the lower-level computational tasks can be done while the data is still in the optical domain. The Nonlinear Schrödinger Kernel computing addresses this predicament. It utilizes optical nonlinearities to map the data onto a new domain in which classification accuracy is enhanced, without increasing the data dimensions.  One limitation of this technique is the fixed optical transfer function, which prevents training and generalizability. Here we show that the optical kernel can be effectively tuned and trained by utilizing digital phase encoding of the femtosecond laser pulse leading to a reduction of the error rate in data classification.


\section*{Introduction}
State-of-the-art deep neural networks with billions of parameters have been deployed for various applications including computer vision, speech processing, engineering design, and scientific discovery. The ever-increasing complexity of the networks is powered by the advances in semiconductor technology described by the Moore’slaw. The projected end of the Moore’s law is a concern for the future evolution of neural networks \cite{jalali2022physics}. Additionally, the neural networks must be trained and this requires a massive amount of labeled data the number of which scales with the network complexity. The unavailability of curated data in many domains precludes the use of deep neural networks. The inference speed of the current neural networks is typically in the millisecond timescale. This is at odds with ultrafast optical sensors which can acquire data at femtosecond time intervals and at picosecond frame intervals. 

This calls for a new look into the approaches that can work in ultrafast timescales with a limited amount of data. Analog computing \cite{solli2015analog} is a promising approach that performs specialized computing in complex nonlinear systems to take over the burden from the digital computer. Photonic hardware accelerator is such an approach that focuses on instantiating complex mathematical models in optical systems. For example, a Convolutional Neural Network (CNN) can be emulated by a stack of spatial light modulators \cite{lin2018all}. A feed-forward neural network can be implemented on a photonic chip using an interconnected photonic circuit \cite{miscuglio2018all}. A photonic reservoir computer can be assembled using an optical amplifier along with a feedback loop \cite{duport2012all}. However, all demonstrated works are either complex in structure or need a significant increase in data dimensionality, which has held back their practical application.

Recently, a technique for performing data classification in the ultrafast timescale, called Nonlinear Schrödinger Kernel has been introduced \cite{zhou2022nonlinear}. It draws an analogy to numerical kernel computing which employs a nonlinear transformation of data without explicitly increasing the data dimensionality. It utilizes optical nonlinearities to map the data onto a new domain in which classification accuracy is enhanced. This is done by modulating the data onto the optical spectrum of a femtosecond pulse and performing complex nonlinear transformation through a nonlinear optical media. Therefore, it was originally developed for applications such as time stretch instruments where in the measurement process, data is encoded onto the spectrum of an ultrashort laser pulse. Previous studies have shown that this technique achieves similar results as a traditional numerical kernel with orders of magnitude lower latency on small datasets. Moreover, it does not increase the dimension of data and hence does not sacrifice the computing resource. One of the limitations of this technique has been the fixed optical transfer function, which prevents training and generalizability, meaning the performance is highly dependent on the dataset.

To solve this problem, we propose an approach to tune and optimize the nonlinear optical transfer function of the kernel. While the nonlinear optical system parameters cannot be tuned, the nonlinear process can be effectively engineered by varying the phase of the input data \cite{chou2005adaptive}. This is based on the insight that most nonlinear interactions, such as self-phase modulation, four-wave mixing, etc. are coherent processes that depend on the input phase. Therefore, tuning the spectral phase of the input data will inevitably change the nonlinear interaction among spectral components. Since the data is encoded onto the spectrum, this process changes the nonlinear interaction among data dimensions.

In this paper, we first investigate the key effect inside the optical kernel and then propose a tuning scheme utilizing complex data encoding. To evaluate the effect of tuning, a digital feedback loop is added to the system to obtain the optimal phase code that gives the best classification accuracy. The effect of phase encoding and optimization is demonstrated on three datasets: time stretch cell image \cite{chen2016deep}, phalanges bones outline \cite{bagnall16bakeoff}, and Electroencephalogram (EEG) \cite{Dua:2019}.

\section*{Experimental System}

\begin{figure}

\centering
\includegraphics[width=75mm]{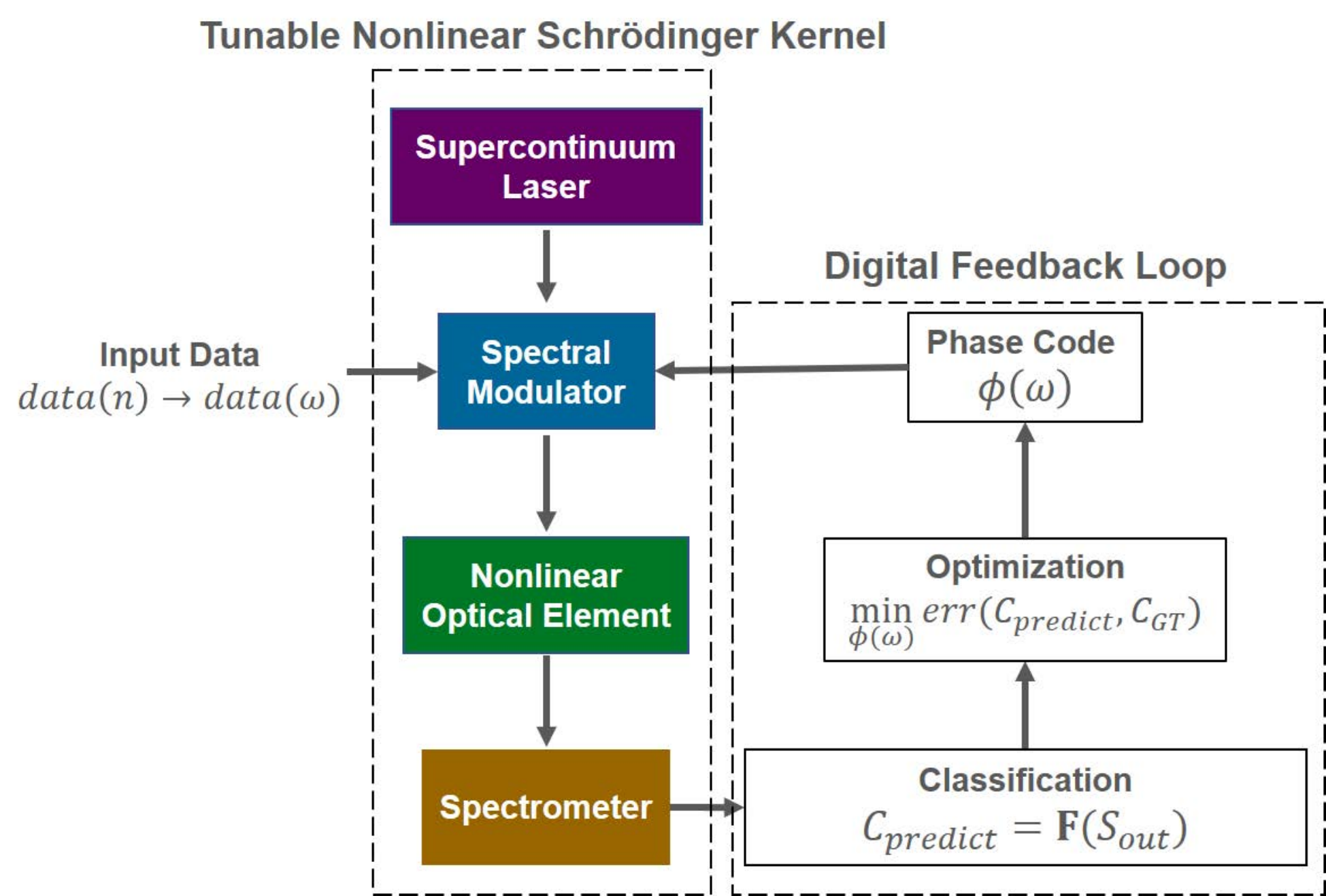}
\caption{Optimization of a tunable Nonlinear Schr{\"o}dinger Kernel. The system contains a tunable Nonlinear Schr{\"o}dinger Kernel and a digital feedback loop. In the tunable Nonlinear Schr{\"o}dinger Kernel, the phase-encoded input $data(n)$ is mapped onto the spectrum of a supercontinuum laser via spectral modulation. The modulated laser propagates through a nonlinear optical element, where the nonlinear process is engineered by the phase code $\varphi(\omega)$. The output spectrum of the nonlinear optical element $S_{out}$ is then acquired using a spectrometer and sent to a classifier F. The classification error is calculated by comparing the predicted class $C_{predict}$ and the ground truth $C_{GT}$. This error is used as an input to an optimization algorithm to update the phase code for achieving lower classification errors. The details can be found in the Materials and Methods.}
\label{fig:NLSKtrain} 
\end{figure}

Fig. \ref{fig:NLSKtrain} shows the system block diagram featuring closed-loop optimization of the optical kernel. This system contains a highly tunable Nonlinear Schr{\"o}dinger Kernel and a digital feedback loop. In the tunable kernel, the input $data(n)$ is mapped into the spectrum domain ($data(\omega)$) and amplitude modulated onto a supercontinuum femtosecond laser pulse. The data then travels through a nonlinear optical element where it is nonlinearly transformed. The transformed data is finally captured by a spectrometer and sent to a backend digital classifier which adopts a light machine learning model. To avoid increasing data dimensions, the output spectrum of the kernel is sampled to match the dimension of the input data. The tuning is achieved by modulating a spectral phase $\varphi(\omega)$ onto the laser pulse during spectral modulation, equivalent to encoding the phase of the input data. This ‘phase code’ modifies the nonlinear interaction in the nonlinear element and hence engineers the optical kernel. In the digital feedback loop, the algorithm compares the predictions with the ground truth and calculates the average classification error for the whole dataset. Subsequently, a new phase code is generated by an optimizer who aims at minimizing this error. The system runs iteratively to obtain the optimal phase code, through which the optical kernel is trained for optimal performance. The details of the experiment system can be found in the Materials and Methods section.

\section*{Results}

The main subject of this paper is the optimization and training of the nonlinear optical kernel. We first describe the critical importance of nonlinearities in this approach. 

\subsection*{The Crucial Role of Optical Nonlinearity} 

\begin {figure}
\centering
\includegraphics[width = \textwidth]{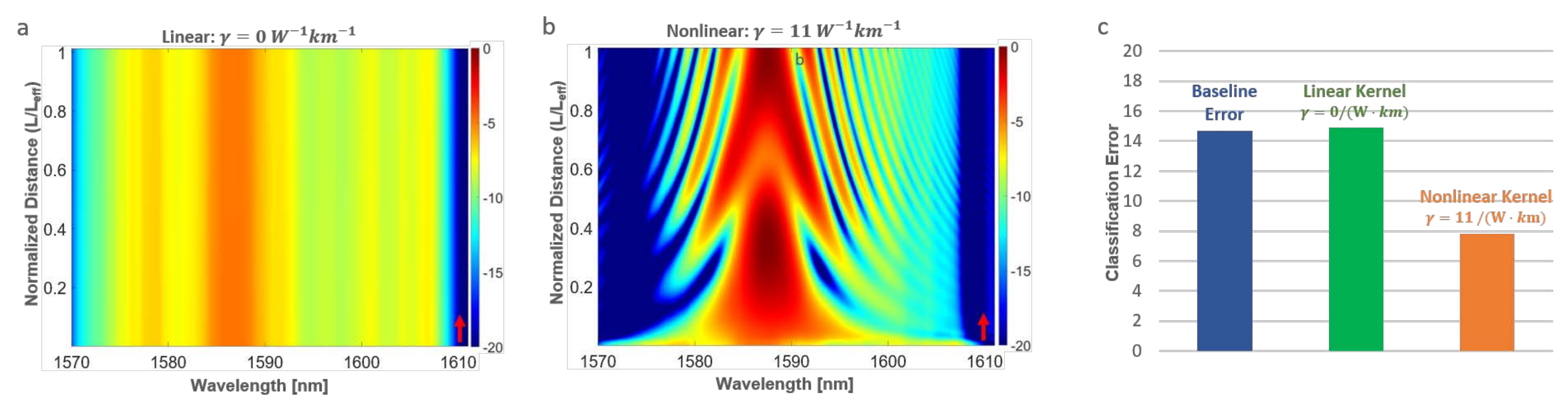}
\caption {(a) The evolution of the optical spectrum in the linear optical kernel where nonlinear coefficient $\gamma=0/(W\cdot km)$(b) The evolution of the optical spectrum in the nonlinear optical kernel where $\gamma=11/(W\cdot km)$. (c) Bar chart comparing the classification error for three cases: the baseline error calculated without kernel (blue, $14.7\%$), the linear optical kernel (green, $14.9\%$), and the nonlinear optical kernel (orange, $7.8\%$). The baseline error is calculated by directly feeding the input data to the digital backend, which in this case is a linear support vector machine (SVM) classifier. For (a) and (b), the horizontal axis is the wavelength, the vertical axis is the propagation distance (normalized to the effective length of the optical element). The color indicates the optical intensity in the log scale with color bar on the side. The red arrow point to the propagation direction.
}
\label {fig:NLSKnl}
\end {figure}

In Fig. \ref{fig:NLSKnl} we show via simulation the evolution of a femtosecond pulse, that has been spectrally modulated with data, through the nonlinear optical element. The output spectrum is classified using a simple machine learning algorithm. Also shown in Fig. \ref{fig:NLSKnl} is the error produced by using a linear support vector machine (SVM) classifier as the digital backend. Here the nonlinear kernel is fixed, i.e. not tuned. As described in the Methods section, the data is the 1-D linescan images of biological cells flowing through a microfluidic channel. The images are captured by a time stretch microscope \cite{chen2016deep}\cite{li2019deep}.  The dataset contains three types of images: (1) no cell is present, (2) a normal cell, and (3) a cancer cell.  The classification task is to distinguish three different types. 

In Fig. \ref{fig:NLSKnl}a, a linear kernel is simulated by setting the nonlinear coefficient to zero. As expected, the optical spectrum remains unchanged. As shown in Fig. Fig. \ref{fig:NLSKnl}c the classification error remains almost unchanged at $14.9\%$ compared to the $14.7\%$ baseline error obtained by feeding the data directly into the backend digital classifier. Fig. Fig. \ref{fig:NLSKnl}b shows the propagation of the same spectrally modulated pulse in the nonlinear element. To avoid an increase in data dimension, we operate in the spectrum narrowing regime. This occurs when the pulse undergoing self-phase modulation has a negative chirp \cite{agrawal2012nonlinear}. The output spectrum is sampled such that it has the same dimensions as the input data (128) as explained in the Methods section. The nonlinear transformation reduces the classification error to $7.8\%$, confirming the utility of the optical kernel in enhancing machine learning without sacrificing (ie. increasing) the data dimensionality.

Comparing the effect of linear and nonlinear kernels, it can be observed that the enhancement in classification accuracy cannot be achieved without optical nonlinearity. However, as a proper machine learning technique, the nonlinearity must be tunable so it can be optimized. Previous research has demonstrated the successful control of optical nonlinearity using the spectral phase modulation of the input light \cite{chou2005adaptive}. Here we apply to same technique to tuning and optimization of the nonlinear optical kernel where the optimization is guided by the classification error. 

\subsection*{Training of Optical Nonlinearities for Machine Learning}

In this section, we demonstrate that the nonlinear optical kernel can be tuned by applying phase encoding to the input data. As mentioned in the introduction, the intuition behind this approach is as follows. Nonlinear optical interactions such as self-phase modulation, Four Wave Mixing (FWM), etc. are coherent in nature, i.e. they are sensitive to the phase of the input pulse. It then follows that manipulating the input phase influences the output produced by the optical nonlinearity.

\begin {figure}
\centering
\includegraphics[width = \textwidth]{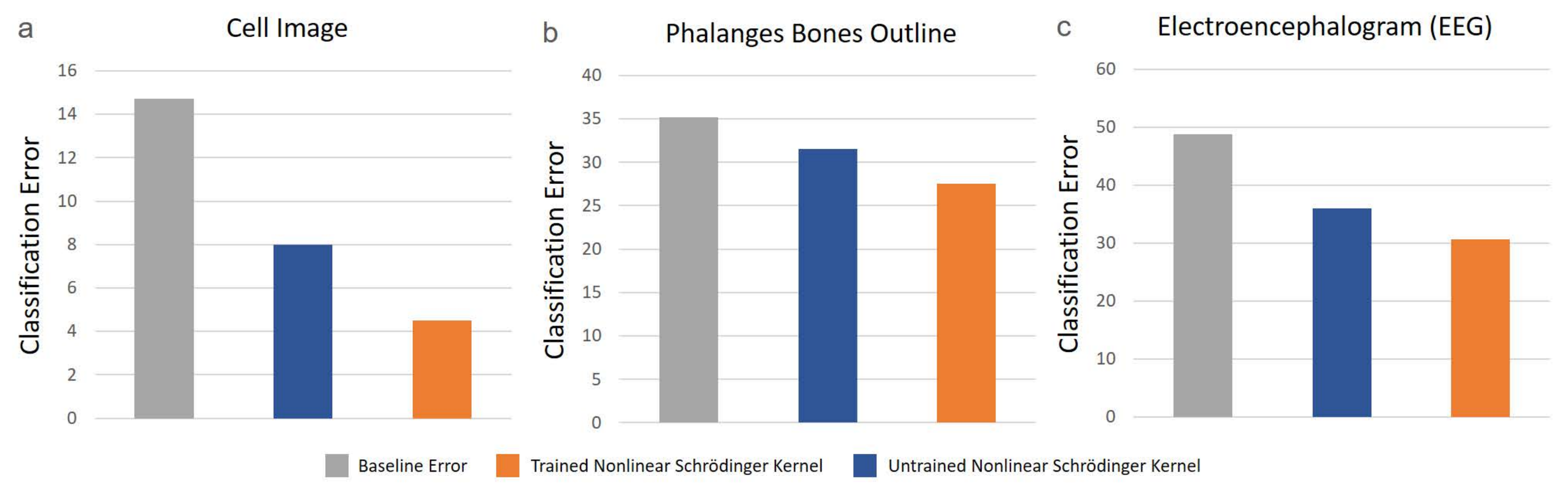}
\caption {Optimization of Nonlinear Schr{\"o}dinger Kernel on three datasets: (a) Time stretch biological cell image (b) Phalanges bones outline (c) Electroencephalogram (EEG). In each bar chart, the classification error for three cases is compared: Baseline error (gray), untrained Nonlinear Schr{\"o}dinger Kernel (blue), and trained Nonlinear Schr{\"o}dinger Kernel (orange). The baseline error is calculated by feeding the input data directly into the digital backend – a linear support vector machine (SVM) classifier. All the results are calculated via 3-fold cross validation.
}
\label {fig:NLSKopt}
\end {figure}

The experimental implementation is shown in Fig. \ref{fig:NLSKtrain}. Spectral phase modulation within a digital feedback loop controls the nonlinear optical interactions and hence tunes the optical kernel. A genetic algorithm arrives at the optimal phase code that minimizes the error of the digital classifier. The results for three datasets are experimentally demonstrated. The datasets described in the Materials and Methods include cell images (Fig. \ref{fig:NLSKopt}a) \cite{chen2016deep}\cite{li2019deep}, phalanges bones outline (Fig. \ref{fig:NLSKopt}b)\cite{bagnall16bakeoff}, and EEG (Fig. \ref{fig:NLSKopt}c) \cite{Dua:2019}.

Fig. \ref{fig:NLSKopt} shows that, in all three datasets, both the trained (orange), as well as the fixed (untrained) kernel (blue), produce a lower error rate compared to the baseline case where no optical kernel is used (gray). However, the trained optical kernel leads to a lower error than the fixed kernel. The baseline error is obtained by feeding the input data directly to the digital backend. The results prove that the phase encoding can effectively tune the performance of the Nonlinear Schr{\"o}dinger Kernel, and combined with the digital feedback loop renders the optical kernel trainable. 

In these experiments, the phase code is generated using a polynomial with two tunable parameters for the second-order and the third-order coefficients. This scheme can be easily expanded to more parameters to provide additional degrees of freedom. Details of the experiments are also in the Materials and Methods.

\section*{Limitations}

Some of the limitations of this technique and potential future research are as follows. First, the maximum allowed dimension of the input data is dependent on spectral modulation. In our experiments, we use a commercial waveshaper (details in the Methods section) with 500 pixels. This limits the maximum dimension of the input data to 500. Second, even though the kernel can be trained, the performance in terms of classification error is still data-dependent, as seen in \ref{fig:NLSKopt}. Such is the case with all machine learning techniques because they are statistical in nature (as opposed to deterministic). One possible direction for future research is to correlate the classification performance with the properties of the input data to identify the type of data for which optical kernel computing is most effective. 

\section*{Conclusion}

The recently introduced optical kernel computing utilizes optical nonlinearities to transform data such that nonlinear classification can be done with a computationally light linear digital classifier. The so-called Nonlinear Schr{\"o}dinger Kernel computing is ideal for low latency classification of data that is modulated onto the spectrum of femtosecond lasers. Such is the case with time stretch imaging and spectroscopy instruments \cite{mahjoubfar2017time}\cite{zhou2022unified}. In the previous implementation of this technique, the property of the kernel was entirely governed by the nonlinear coefficient of the optical medium. Hence the kernel could not be trained or optimized as required in machine learning tasks. In this chapter, we presented a solution to this predicament by introducing spectral phase modulation of the input pulse within a digital feedback loop. Phase modulation influences how data is transformed by nonlinear optical interactions and allows the optical kernel to be trained. The training is shown to reduce the classification error on three diverse datasets.

\section*{Materials and Methods}

In this section, we provide details of (1) experimental implementation for kernel optimization, (2)  simulation study showing the critical role of optical nonlinearities, (3) mathematical formulation of the optical kernel, and (4) the datasets and machine learning model.

\subsection*{Experiments}

The physical implementation of the closed loop optical kernel computing system follows Fig. \ref{fig:NLSKtrain}. The supercontinuum laser is a mode-locked Erbium-doped fiber laser (ELMO) followed by an Erbium-doped fiber amplifier (ELMA), both from Menlo Systems. It produces $< 90 fs$ laser pulses centered at $1560 nm$ with $90 nm$ bandwidth. The approximate pulse peak power is $30$ W after the amplifier. The data modulates the optical spectrum using a Finisar (now II-VI / Coherent) Waveshaper model 1000 S/L. It operates in the L band ($1567 nm$ ~ $1609 nm$) with a 500-pixel resolution. The Waveshaper performs both amplitude and phase modulation simultaneousl. The amplitude is the input data (scaled between 0 to 1), while the phase is the (phase) code that is applied to tune the nonlinear effects. The nonlinear optical element is a $500 m$ highly nonlinear fiber (HNLF) from Corning. It has a nonlinear coefficient $\gamma =11/(W\cdot km)$, a low dispersion of $D=1.43 ps/(nm\cdot km)$, and dispersion slope $S=0.04 ps/(nm^2\cdot km)$ at $1588 nm$. An Ando AQ6317B optical spectrum analyzer with $0.1 nm$ resolution measures the output spectrum. The sampling range of the spectrum analyzer is set differently for each dataset depending on the actual bandwidth of the output. For the cell image and phalanges bones outline dataset, it is $1545 nm$ ~ $1635 nm$. For the EEG dataset, it is $1550 nm$ ~ $1630 nm$. The measured spectrum is resampled so that it has the same dimension as the input. It is then min-max standardized and sent to a linear SVM (support vector machine) classifier.

The digital feedback loop is implemented in MATLAB on a computer with 16 GB RAM memory. It runs iteratively to optimize the performance of the Nonlinear Schr{\"o}dinger Kernel. In each iteration, the classification error rate is the average over the entire dataset and is calculated via 3-fold cross validation. It is sent to the optimizer which is a genetic algorithm \cite{Holland:1975} from MATLAB Global Optimization Toolbox. To minimize the classification error, it compares the result from the current iteration with the previous ones and generates the phase code for the next iteration. In this paper, the phase code is calculated by a third-order polynomial with adjustable coefficients,
\begin{equation}
    \varphi(\omega) = a\omega^2+b\omega^3
\end{equation}
here, a and b are the tunable coefficients generated by the optimizer. The first-order coefficient is not taken into consideration as it simply causes a constant delay and does not affect the nonlinear optical process. The optimization results are shown in Fig. \ref{fig:NLSKopt}. The optimal coefficients for the cell image dataset (Fig. \ref{fig:NLSKopt}a) are shown in Table \ref{tabopt}. The optimized classification error, $Err_{trained}$ are also attached in comparison with the unoptimized classification error $Err_{untrained}$ and the baseline error $Err_{baseline}$.

\begin{table}
\centering
\begin{tabular}{||c c c c||} 
 \hline
   & Cell Imgage & Phalanges Bones & EEG \\ [0.5ex] 
 \hline
 $a(\times 10^{-24})$ & 1.353 & 0.854 & 1.022 \\ 
 \hline
 $b(\times 10^{-37})$ & -3.628 & 0.925 & -4.742 \\
 \hline
 $Err_{trained}$ & $4.5\%$ & $27.5\%$ & $30.7\%$ \\
 \hline
 $Err_{untrained}$ & $8.0\%$ & $31.5\%$ & $36.04\%$ \\
 \hline
 $Err_{baseline}$ & $15.0\%$ & $35.2\%$ & $30.7\%$ \\ 
 \hline
\end{tabular}
\caption{Optimal Phase Code}
\label{tabopt}
\end{table}

\subsection*{Insight into the Critical Role of Optical Nonlinearity}

To assess the role of optical nonlinearities in the operation of the Nonlinear Schr{\"o}dinger Kernel, a computer model is created in MATLAB. The supercontinuum laser source is modeled as a transform-limited pulse with a supper-Gaussian spectrum centered at $1588 nm$. It has a $40 nm$ bandwidth consistent with the passband of the spectral modulator (waveshaper). The spectral modulation is modeled by multiplying the input data (time stretch cell image data, scaled between 0 to 1) by the laser spectrum. For these simulations, the phase code is set to 0 to simulate the unoptimized (open loop) system. The modulated laser pulse is then sent through a nonlinear optical element, in this case, an HNLF. The complex propagation in the fiber is modeled by solving the time domain Nonlinear Schrödinger Equation (NLSE) using the split-step Fourier method (SSFM) implemented in MATLAB. This algorithm divides the fiber into short segments (steps) to separate different effects, computing iteratively to obtain an approximate solution \cite{agrawal2012nonlinear}. In each step, the spectrum of the optical pulse is recorded to track the evolution of the input, as shown in Figure 2. The output of the HNLF is measured using a spectrometer, which is modeled by a fast Fourier transform (FFT) and absolute square.  Finally, the collected spectrum is resampled so that it has the same dimension as the input, mean-std standardized, and sent to a linear SVM classifier.

For the simulation shown in Fig. \ref{fig:NLSKnl}, the length of the fiber is set to $500 m$. The optical loss is $0.02 dB/km$, the dispersion coefficient (D) is $1.43 ps/(nm\cdot km$), the dispersion slope (S) is $0.04 ps/(nm^2\cdot km)$ at $1588 nm$, and the optical power $P$ is $20 W$. For the linear kernel, the nonlinear coefficient ($\gamma$) is $0 rad/(W\cdot km)$. While for the nonlinear kernel, $\gamma=11 rad/(W\cdot km)$. The spectrum is sampled from $1570 nm$ to $1610 nm$ with 128 sampling points to match the dimension of the input data. 

The MATLAB model runs on a server equipped with 64 GB memory and an NVIDIA RTX TITAN GPU with 24 GB memory. The SSFM algorithm is modified from the open-source function ssprop and accelerated by implementation on GPU using CUDA.  

\subsection*{Mathematical Model}

The mathematical description of the Nonlinear Schrödinger Kernel computing is provided in this section. As shown in \ref{fig:NLSKtrain}, the data is first mapped into the optical spectrum by assigning each entry of the data to a frequency. In the current experiments, the data resides in a file and is modulated onto the femtosecond pulse spectrum using a waveshaper. The waveshaper performs the following mapping $M$:
\begin{equation}
    M: data(n) \rightarrow data(\omega)
\end{equation}
$data(n)$ is the input where n is its $n$-th entry, and $\omega$ is the corresponding optical frequency. This filter is then applied to a supercontinuum laser pulse:
\begin{equation}
    E_{in}(\omega) = \sqrt{data(\omega)}\cdot E_{laser}e^{\varphi(\omega)}
\end{equation}
here, $E_{laser}$ is the supercontinuum laser source, $\varphi(\omega)$ is the phase code which is also applied by the waveshaper for the purpose of training. $E_{in}$ is the modulated pulse which will then enter the nonlinear stage. The data is modulated onto the power spectrum, which is equivalent to multiplying its square root by the E-field. The modulated laser is sent into a nonlinear optical element, where complex nonlinear transformation is governed by the NLSE:
\begin{equation}
    \frac{\partial E(t)}{\partial z} = (\hat{D} + \hat{N})E(t)
\end{equation}
here $E$ is the electric field of the optical pulse, z is the propagation distance, $\hat{D}$  is the dispersion operator, and $\hat{N}$ is the nonlinear operator, which can be calculated from the parameters of the nonlinear optical element [11]. By solving NLSE using the initial condition:
\begin{equation}
    E_in{t} = \frac{1}{2\pi}\int_{-\infty}^{+\infty}e^{i\omega t}E_{in}(\omega)d\omega
\end{equation}

The output $E_{out}(t)$ is thus obtained. The nonlinearly trasnformed input data is subsequently acquired along with the laser spectrum $S_{out}(\omega)$. According to the Fourier transform, the spectrum is:
\begin{equation}
    E_{out}(\omega) = \int_{-\infty}^{+\infty}e^{-i\omega t} E_out(t)dt
\end{equation}
\begin{equation}
    S_{out}(\omega) = |E_{out}(\omega)|^2
\end{equation}

The classification is then performed on the measured output spectrum:
\begin{equation}
    C_{predict} = F(S_{out})
\end{equation}
here, $C_{redict}$ is the predicted class, and $F$ is a linear SVM classifier. We chose the linear SVM for its ubiquity and simplicity. Other classifiers can also be used. 

In the feedback loop, the classification error $err(C_{predict},C_{GT})$ is calculated by comparing the prediction and the ground truth. Since $C_{predic}$ is a function of phase code $\varphi(\omega)$, we can obtain the optimal $\varphi(\omega)$ by minimizing the error:
\begin{equation}
    \min_{\varphi(\omega)} err(C_{predict},C_{GT})
\end{equation}

In this paper, the optimizer is a genetic algorithm, as mentioned in the experiments section.

\subsection*{Datasets and Machine Learning Model}

The performance of the trained Nonlinear Schrödinger Kernel is evaluated with three datasets: time stretch cell image \cite{chen2016deep}\cite{li2019deep}, phalanges bones outline \cite{bagnall16bakeoff}, and EEG \cite{Dua:2019}. The details are provided as follows.

The time stretch cell image dataset is the image of cells acquired in time stretch microscopic flow cytometry \cite{chen2016deep}\cite{mahjoubfar2017time}. In this setup, a femtosecond laser pulse is spatially dispersed into a collimated 1-D rainbow using a pair of diffraction gratings. The rainbow illuminates a microfluidic channel while sample blood cells pass by. Through this, the spatial features of the illuminated cells are modulated onto the spectrum of the femtosecond laser pulse. Those features are then read out in realtime using a time stretch spectrometer, where a low-loss dispersive fiber maps the spectrum in time. The temporal waveforms that mimic the optical spectrum are digitized by a realtime analog to digital converter. Each frame is one linescan of the cells moving along with the microfluidic flow. The details of the setup are described in our earlier publications \cite{mahjoubfar2017time}. The dataset contains three types of linescans: the background, the colon cancer blood cells (SW-480 epithelial), and the white blood cells (OT-II hybridoma 497 T-cells). Each class includes 200 observations (waveforms), and each waveform has 128 dimensions (spatial features). The details can also found in

Phalanges bones outline is an open-source dataset from the UCR Time Series Classification Archive \cite{bagnall16bakeoff}. It studies the correctness of an automatic phalanges bone outline extraction algorithm. The algorithm is applied to X-ray images for extracting the outlines of three bones of the middle finger (phalanges). The extraction is then evaluated by three human evaluators. This paper uses a selection of 400 observations evenly distributed in two classes: correct outlines and incorrect outlines. The details of the dataset can be found in \cite{uea45085}.

The EEG dataset is an open-source dataset from UCI Machine Learning Repository \cite{Dua:2019}. It comes from the test that measures the brain’s electrical activity. The data is collected using the Emotiv EEG Neuroheadset for studying the correspondence between EEG signals and human eye motions. This paper uses a subsample of 400 observations from the dataset with two eye motion states (200 for each): eye open and eye closure. Each observation has 14 dimensions. 

The backend digital classifier in this study adopts a linear SVM model \cite{vapnik1963pattern}. The classification errors are calculated via three-fold cross validation.

\section*{Acknowledgement}

This project was supported by DARPA-MTO PEACH program under contract number HR00111990050. BJ conceived the idea and the technical approach. TZ performed the experiments and data analytics. BJ and TZ wrote the manuscript.

\bibliography{library}

\bibliographystyle{abbrv}

\end{document}